\documentclass[conference]{IEEEtran}
\usepackage{amsmath,amssymb}
\usepackage{bm}
\usepackage{graphicx}
\usepackage{cite}
\usepackage{algorithmic}
\usepackage{algorithm}
\usepackage{booktabs}
\usepackage[T1]{fontenc}
\usepackage[utf8]{inputenc}
\graphicspath{{figures/}{globecom_python/figures/}{./}}

\newcommand{\tr}{\operatorname{tr}}
\newcommand{\diag}{\operatorname{diag}}
\newcommand{\Real}{\operatorname{Re}}
\newcommand{\vect}[1]{\mathbf{#1}}
\newcommand{\mat}[1]{\mathbf{#1}}
\newcommand{\FIM}{\mathbf{I}}

\begin{document}

\title{Localization Bounds and Shape Design for Cell-Free ISAC with Fluid Intelligent Metasurfaces}

\author{
\IEEEauthorblockN{Changhao He\IEEEauthorrefmark{1},
Jiguang He\IEEEauthorrefmark{2},
Asmaa Abdallah\IEEEauthorrefmark{1},
Ahmed M.~Eltawil\IEEEauthorrefmark{1}}
\IEEEauthorblockA{
\IEEEauthorrefmark{1}King Abdullah University of Science and Technology (KAUST), Saudi Arabia\\
\IEEEauthorrefmark{2}Great Bay University (GBU), China\\
{\small\{changhao.he, asmaa.abdallah, ahmed.eltawil\}@kaust.edu.sa; jiguang.he@gbu.edu.cn}}
}

\maketitle

\begin{abstract}
Fluid intelligent metasurfaces (FIMs) reshape antenna geometry to improve angular sensitivity. This paper studies their role in cell-free integrated sensing and communication, where distributed access points provide distinct observation directions. Under calibrated path gains, we derive the conditional angular Fisher information matrix, retain cross-AP coupling, and map it to a position-domain Cram\'er--Rao bound (CRB). A diagonal-information specialization exposes the dependence on angular diversity, while a multi-target extension quantifies the loss from jointly estimating other target positions. We use CRB-driven coordinate search to design transmit and receive shapes and evaluate sensing performance, equal-resource AP scaling, and MRT-based communication--sensing allocation. In the considered four-Tx/four-Rx configuration, shape design reduces the position root CRB from 0.300 to 0.227 m at 10 dB reference SNR. Two-target results further illustrate how cross-target coupling increases the joint localization bound at small separations.
\end{abstract}

\section{Introduction}
Integrated sensing and communication (ISAC) uses shared infrastructure for data delivery and environmental sensing~\cite{liu2022isac,liu2020crb_dfrc}. Cell-free networks extend this principle through geographically distributed access points (APs), providing multiple sensing directions~\cite{bjornson2020cellfree_book,demirhan2024cellfree_isac,behdad2024multistatic}. Distributed observations can improve localization geometry, but their value also depends on the angular sensitivity of each local array.

Fluid intelligent metasurfaces (FIMs) adapt array geometry by moving radiating elements~\cite{zhu2024modeling}. Related fluid- and movable-antenna architectures have been studied for communication and sensing~\cite{wong2020fluid,zhu2024movable,ma2024movable_sensing}. Unlike phase-only control, physical displacement changes the steering-vector derivative and can improve sensitivity near the endfire direction of a fixed linear array. This benefit depends on the observation angle and need not be uniform across a distributed network.

Recent FIM-enabled ISAC work addresses CRB-driven beamforming and shape design~\cite{zhang2025fimisac} as well as FIM deformation and user-fairness optimization~\cite{huang2026fimisac}. In a distributed system, waveform correlation couples the information associated with different transmitting APs. Moreover, improved angular sensitivity does not translate directly into the same localization gain: the AP geometry and the full information matrix must be considered together.

We derive a conditional information model for FIM-assisted cell-free ISAC with known, calibrated path gains. The analysis retains cross-AP and transmit--receive information coupling, provides a geometric closed form under diagonal angular information, and extends to jointly estimated target positions. We then apply a feasible coordinate-search method to shape design. Numerical results distinguish angular accuracy from localization accuracy, compare AP layouts under fixed total resources, and examine MRT power allocation and two-target information coupling. The scope is local CRB analysis rather than detection, association, or unknown-gain estimation.

\section{System Model}
\label{sec:system}

\begin{figure}[t]
\centering
\includegraphics[width=\columnwidth]{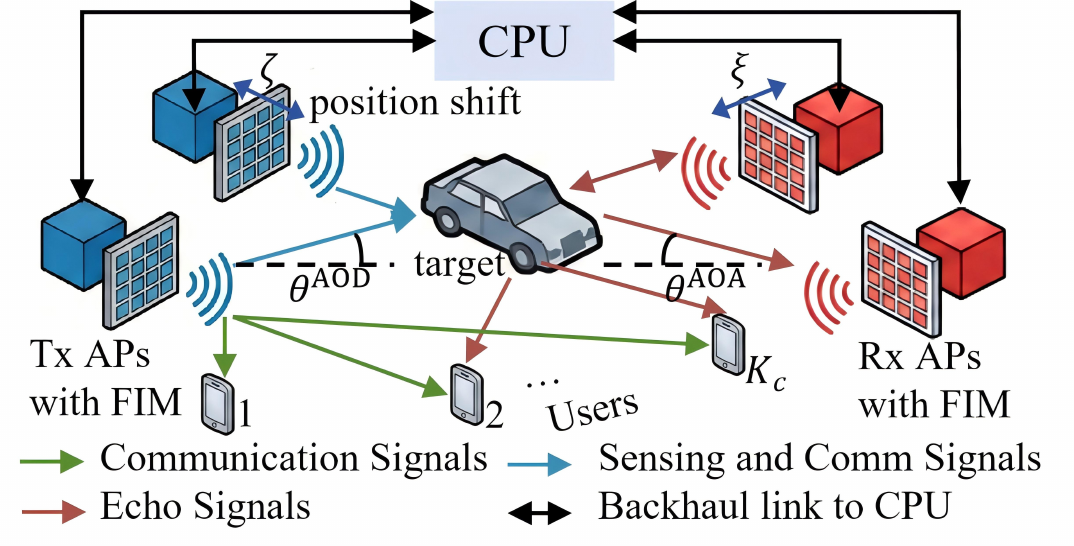}
\caption{FIM-assisted cell-free ISAC. Distributed Tx APs probe the target and serve users, Rx APs receive echoes, and the CPU coordinates processing. Displacing the Tx/Rx FIM elements reconfigures their array geometry.}
\label{fig:sysmodel}
\end{figure}

We consider a cell-free ISAC system consisting of
$M_t$ transmitting access points (Tx APs) and
$M_r$ receiving access points (Rx APs) distributed
over a geographical area and coordinated by a central
processing unit (CPU), as illustrated in Fig.~\ref{fig:sysmodel}.
Each AP is connected to the CPU via a reliable
backhaul link, enabling cooperative signal processing
and joint sensing operations. The positions of the Tx and Rx APs are denoted by $\mathbf{p}^{(\mathrm{Tx})}_{m_t} \in \mathbb{R}^2$ and $\mathbf{p}^{(\mathrm{Rx})}_{m_r} \in \mathbb{R}^2$, respectively, where $m_t = 1,\ldots,M_t$ and $m_r = 1,\ldots,M_r$.

Each Tx AP is equipped with an $N_t$-element one-dimensional (1D) FIM array,
and each Rx AP with an $N_r$-element 1D FIM array.
The positions of FIM elements at each AP can be
independently adjusted within a bounded region,
as detailed in Section~\ref{sec:FIM_morph}.

{The $M_t$ Tx APs simultaneously serve $K_c$ single-antenna downlink users while, together with the $M_r$ Rx APs, also sense a point target located at $\mathbf{p}_0 = [x_0, y_0]^\top$.}

\subsection{Sensing Model}
Due to the distributed AP deployment, each AP observes the target from a different spatial direction. The resulting angle of departure (AoD) from Tx AP $m_t$ and the angle of arrival (AoA) at Rx AP $m_r$ are given by
\begin{equation}
\theta_{m_t}^{(\text{AoD})} = \angle(\vect{p}_0 - \vect{p}_{m_t}^{(\text{Tx})}), \quad
\theta_{m_r}^{(\text{AoA})} = \angle(\vect{p}_0 - \vect{p}_{m_r}^{(\text{Rx})}).
\label{eq:angles}
\end{equation}
The distributed geometry provides $M_t+M_r$ angle parameters whose information can be coupled; distinct AP positions need not produce distinct target bearings.

\subsubsection{FIM-Augmented Array Response}\label{sec:FIM_morph}

Let $\boldsymbol{\zeta}_{m_t} = [\zeta_{m_t,1}, \ldots, \zeta_{m_t,N_t}]^\top$ denote the displacement vector of the FIM elements at the transmit AP $m_t$, where each element displacement is bounded as $|\zeta_{m_t,n}| \le \tilde{\zeta}$ for $n = 1, \ldots, N_t$. The morphing range $\tilde{\zeta}$ typically lies between $0.5\lambda$ and $2\lambda$~\cite{zhu2024modeling,ma2024movable_sensing}, corresponding to millimeter-scale adjustments at mmWave frequencies. Similarly, for the receive AP $m_r$, the displacement vector is defined as $\boldsymbol{\xi}_{m_r} = [\xi_{m_r,1}, \ldots, \xi_{m_r,N_r}]^\top$, with each displacement satisfying $|\xi_{m_r,n}| \le \tilde{\xi}$.

The FIM array response at Tx AP $m_t$, assuming half-wavelength element spacing, is $[\vect{a}_{m_t}]_n = N_t^{-1/2}e^{j\phi_{m_t,n}}$, with $\vect{a}_{m_t}\in\mathbb{C}^{N_t}$, where the phase of the $n$-th element is
\cite{zhu2024modeling}
\begin{equation}
\phi_{m_t,n} = (n\!-\!1)\pi\sin\theta_{m_t} + \tfrac{2\pi}{\lambda}\,\zeta_{m_t,n}\cos\theta_{m_t}.
\label{eq:phase}
\end{equation}
The first term corresponds to the standard half-wavelength ULA,
while the latter captures FIM's physical shape deformation;
setting $\bm{\zeta}_{m_t} = \vect{0}$ recovers the conventional
ULA response.
The Rx AP array response $\vect{a}_{m_r}$ follows the same
structure with $m_t$, $N_t$, and $\boldsymbol{\zeta}_{m_t}$  replaced by $m_r$, $N_r$, and  $\boldsymbol{\xi}_{m_r}$, respectively.

To characterize angular sensitivity, we define the
phase-rate $[\vect{d}_m]_n \triangleq \partial\phi_{m,n}/\partial\theta_m$
for a generic AP $m\in\{m_t,m_r\}$\cite{vanTrees2002detection}:

\begin{equation}
[\mathbf{d}_{m}]_{n} = (n-1)\pi \cos \theta_{m} - \frac{2\pi}{\lambda}\,\zeta_{m,n}\sin \theta_{m},
\label{eq:phase_rate}
\end{equation}
so that $\dot{\mathbf{a}}_{m} \triangleq \partial \mathbf{a}_{m}/\partial\theta_{m} = j\,\mathrm{diag}(\mathbf{d}_{m})\,\mathbf{a}_{m}$.
The quantity {$|[\vect{d}_m]_n|$} measures n$^{th}$ element angular sensitivity;
FIM morphing modifies this via the $\zeta_{m,n}\sin\theta_m$ term.

\subsubsection{Bistatic Sensing Channels}

Since the Tx and Rx APs are spatially separated, the system operates in multi-bistatic configuration. The sensing channel from Tx AP~$m_t$ to Rx AP~$m_r$ via the target is
\begin{equation}
\mat{G}_{m_t,m_r} = \alpha_{m_t,m_r}\,
\vect{a}_{m_r}(\theta_{m_r},\bm{\xi}_{m_r})\,
\vect{a}_{m_t}^H(\theta_{m_t},\bm{\zeta}_{m_t}),
\end{equation}
where $\alpha_{m_t,m_r}$ is the complex path gain. The following analysis conditions on known, calibrated path gains, held fixed when differentiating with respect to position or angle. It therefore gives an angle-based conditional CRB; unknown gains require nuisance-parameter elimination and generally yield a larger bound. Angles are measured from the global $x$-axis, with nominal arrays parallel to the $y$-axis and displacements along the $x$-axis. The CPU knows the transmitted waveforms and coherently aligns the sensing observations.
The received sensing signal at Rx AP~$m_r$ is $\mathbf{Y}_{m_r} = \sum_{m_t=1}^{M_t} \mathbf{G}_{m_t,m_r} \mathbf{X}_{m_t} + \mathbf{N}_{m_r}$, where $\mathbf{X}_{m_t} \in \mathbb{C}^{N_t \times L}$
denotes the transmitted waveform matrix from Tx AP $m_t$,
where $L$ is the number of snapshots.
Each column of $\mathbf{N}_{m_r}$ is distributed as $\mathcal{CN}(\vect{0},\varsigma_{m_r}^2\mathbf{I}_{N_r})$, independently across snapshots and Rx APs.

{\subsection{Communication Model}}

Let $\vect{h}_{k,m_t}\!\in\!\mathbb{C}^{N_t}$ denote the
downlink channel from Tx AP~$m_t$ to user~$k$,
which incorporates the FIM-augmented array response at
AP~$m_t$.
Each Tx AP applies a per-user beamforming vector
$\vect{w}_{k,m_t}\!\in\!\mathbb{C}^{N_t}$, so that the
aggregate transmit signal at AP~$m_t$ is
$\vect{x}_{m_t} = \sum_{k=1}^{K_c}\vect{w}_{k,m_t}s_k + \vect{x}_{m_t}^{(\text{s})}$,
where $s_k$ is the data symbol for user~$k$ with
$\mathbb{E}[|s_k|^2]=1$, and {
$\vect{x}_{m_t}^{(\text{s})}$ is the sensing waveform component.}
The received signal at user~$k$ is
{$y_k = \sum_{m_t=1}^{M_t}\vect{h}_{k,m_t}^H\vect{w}_{k,m_t}\,s_k
+ \sum_{j\neq k}\sum_{m_t=1}^{M_t}\vect{h}_{k,m_t}^H\vect{w}_{j,m_t}\,s_j
\quad + \sum_{m_t=1}^{M_t}\vect{h}_{k,m_t}^H\vect{x}_{m_t}^{(\text{s})} + n_k,$}
where $n_k\sim\mathcal{CN}(0,\sigma_c^2)$ is receiver noise.
{Let $\vect{R}_{m_t}^{(\text{s})} = \mathbb{E}[\vect{x}_{m_t}^{(\text{s})}\vect{x}_{m_t}^{(\text{s})H}]$ denote the sensing waveform covariance at Tx AP $m_t$. Define $\mathsf{S}_k = \bigl|\sum_{m_t}\vect{h}_{k,m_t}^H\vect{w}_{k,m_t}\bigr|^2$,
$\mathsf{I}_k = \sum_{j\neq k}\bigl|\sum_{m_t}\vect{h}_{k,m_t}^H\vect{w}_{j,m_t}\bigr|^2
+ \sum_{m_t}\vect{h}_{k,m_t}^H\vect{R}_{m_t}^{(\text{s})}\vect{h}_{k,m_t}$,
assuming mutually uncorrelated sensing signals across APs and independence from the data symbols, the communication SINR at user~$k$ is $\text{SINR}_k = \mathsf{S}_k \big/ \bigl(\mathsf{I}_k + \sigma_c^2\bigr).$

}

\section{CRB Analysis}
Building on the signal model, we now derive the Fisher information matrix and the resulting CRB for joint angle estimation and target localization.
\label{sec:crb}

The full angle parameter vector is
$\bm{\theta} = [\theta_1^{(\text{AoD})}, \ldots, \theta_{M_t}^{(\text{AoD})}, \theta_1^{(\text{AoA})}, \ldots, \theta_{M_r}^{(\text{AoA})}]^\top \in \mathbb{R}^K$,
where $K = M_t + M_r$. The CRB for estimating $\boldsymbol{\theta}$ is governed by the Fisher information matrix derived below.
\subsection{Fisher Information Matrix}
Applying the Slepian--Bangs formula~\cite{kay1993estimation,vanTrees2002detection} to the received signal model, the $(i,j)$-th element of the Fisher information matrix $\FIM(\bm{\theta})\in \mathbb{R}^{K \times K}$  is
\begin{equation}
    [\FIM(\bm{\theta})]_{ij} = \sum_{m_r} \frac{2}{\varsigma_{m_r}^2} \Real\bigg\{\tr\bigg[\frac{\partial\mat{S}_{m_r}^H}{\partial\theta_j}\frac{\partial\mat{S}_{m_r}}{\partial\theta_i}\bigg]\bigg\},
\label{eq:slepian_bangs}
\end{equation}

where $\mat{S}_{m_r} = \sum_{m_t}\alpha_{m_t,m_r}\vect{a}_{m_r}\vect{a}_{m_t}^H\mat{X}_{m_t} \in \mathbb{C}^{N_r \times L}$ is the noise-free received signal, and $\mat{X}_{m_t} \in \mathbb{C}^{N_t \times L}$ is the transmitted waveform matrix with $L$ denoting the number of snapshots.

To evaluate~\eqref{eq:slepian_bangs}, we compute the partial
derivatives of $\mat{S}_{m_r}$ with respect to the angle
parameters. Differentiating with respect to an AoD parameter
$\theta_{m_t}$ yields
$\partial\mat{S}_{m_r}/\partial\theta_{m_t} =
\alpha_{m_t,m_r}\vect{a}_{m_r}\dot{\vect{a}}_{m_t}^H\mat{X}_{m_t}$,
while differentiating with respect to an AoA parameter
$\theta_{m_r}$ gives
$\partial\mat{S}_{m_r}/\partial\theta_{m_r} =
\sum_{m_t}\alpha_{m_t,m_r}\dot{\vect{a}}_{m_r}
\vect{a}_{m_t}^H\mat{X}_{m_t}$.
Substituting into ~\eqref{eq:slepian_bangs} and grouping
by parameter type, the FIM decomposes into $\FIM(\bm{\theta}) = \bigl[\begin{smallmatrix} \FIM^{(\text{dd})} & \FIM^{(\text{da})} \\ \FIM^{(\text{da})\top} & \FIM^{(\text{aa})} \end{smallmatrix}\bigr]$,where $\FIM^{(\text{dd})}$, $\FIM^{(\text{aa})}$, $\FIM^{(\text{da})}$ are defined in the sequel.

\textbf{AoD--AoD block} $\FIM^{(\text{dd})} \in \mathbb{R}^{M_t \times M_t}$:
\begin{equation}
\FIM^{(\text{dd})}_{m_t,m_t'} = \sum_{m_r=1}^{M_r} \frac{2}{\varsigma_{m_r}^2} \Real\big\{\alpha_{m_t,m_r}\alpha_{m_t',m_r}^*\, \mathcal{Q}_{m_t,m_t'}\big\},
\label{eq:fim_dd}
\end{equation}
where $m_t,m_t'\in\{1,\ldots,M_t\}$ index the Tx APs,
$\mathcal{Q}_{m_t,m_t'} = \mathbf{a}_{m_t}^H \mathbf{D}_{m_t} \mathbf{R}_{m_t,m_t'} \mathbf{D}_{m_t'} \mathbf{a}_{m_t'}$,
$\mathbf{D}_{m_t} = \mathrm{diag}(\mathbf{d}_{m_t})$,
and $\mathbf{R}_{m_t,m_t'} = \mathbf{X}_{m_t}\mathbf{X}_{m_t'}^H$.

For $m_t \neq m_t'$, the off-diagonal terms $\FIM^{(\text{dd})}_{m_t,m_t'}$ represent \textbf{cross-AP Fisher information coupling}, absent in single-AP systems ($M_t = 1$) where $\FIM^{(\text{dd})}$ reduces to a scalar. These terms arise because the same Rx AP simultaneously observes target reflections from multiple Tx APs; non-zero waveform cross-correlation ($\mat{R}_{m_t,m_t'} \neq \vect{0}$) couples their angular information, with the coupling strength governed by $\alpha_{m_t,m_r}\alpha_{m_t',m_r}^*$.

\textbf{AoA--AoA block} $\FIM^{(\text{aa})} \in \mathbb{R}^{M_r \times M_r}$ is always \emph{diagonal}:
\begin{equation}
\FIM^{(\text{aa})}_{m_r,m_r} = \frac{2}{\varsigma_{m_r}^2}\, \gamma_{m_r}(\bm{\xi}_{m_r})\, \beta_{m_r}(\bm{\zeta}_{m_t}),
\label{eq:fim_aa}
\end{equation}
where $\gamma_{m_r}(\bm{\xi}_{m_r}) = \mathbf{a}_{m_r}^H \mathbf{D}_{m_r}^2 \mathbf{a}_{m_r}$ is the Rx effective aperture (depends only on Rx FIM shape $\bm{\xi}_{m_r}$), and $\beta_{m_r}(\bm{\zeta}_{m_t}) = \sum_{m_t}\sum_{m_t'} \alpha_{m_t,m_r} \alpha_{m_t',m_r}^* \mathbf{a}_{m_t}^H \mathbf{R}_{m_t,m_t'} \mathbf{a}_{m_t'}$ is the coherent beam gain,
which depends on all Tx FIM shapes.

AoA estimation accuracy at each Rx AP involves a \textbf{Tx--Rx FIM coupling}: Tx shapes affect AoA information through $\beta_{m_r}$, which can scale as $M_t^2$ under equal-strength phase-aligned contributions and fixed per-AP waveform energy.
This multiplicative structure couples Tx shape design to AoA information at all Rx APs; improving one link need not improve every link.
The effective Rx aperture $\gamma_{m_r}$ depends \emph{solely} on the local Rx FIM shape $\bm{\xi}_{m_r}$, but the full localization objective also depends on the AoD--AoA cross block.

This separation between local aperture and network-dependent coherent
gain is useful for interpreting the shape variables. Receive displacements
act directly on the receiver's phase-rate vector, whereas transmit
displacements also influence the total coherent signal incident on that
receiver. Therefore a transmit-shape change can simultaneously alter
several receive-angle information terms. Optimizing a sum of independent
single-AP angular CRBs would ignore these network interactions and need
not minimize the position-domain objective.

\textbf{AoD--AoA cross block} $\FIM^{(\text{da})} \in \mathbb{R}^{M_t \times M_r}$ is defined as:
\begin{align}
\FIM^{(\text{da})}_{m_t',m_r'} = \tfrac{2}{\varsigma_{m_r'}^2}\Real\bigg\{
-\mu_{m_r'}\sum_{m_t}\alpha_{m_t,m_r'}\alpha_{m_t',m_r'}^* \notag\\
\times\, \vect{a}_{m_t}^H\mat{R}_{m_t,m_t'}\mat{D}_{m_t'}\vect{a}_{m_t'}\bigg\},
\label{eq:fim_da}
\end{align}\normalsize
where $\mu_{m_r} = \vect{a}_{m_r}^H\mat{D}_{m_r}\vect{a}_{m_r}$ is the Rx directional gain.

\subsection{Localization CRB via Jacobian}

To obtain the position-domain CRB, we apply the transformation $\FIM(\vect{p}_0) = \mat{J}^\top\FIM(\bm{\theta})\mat{J}$, where the Jacobian $\mat{J} \in \mathbb{R}^{K \times 2}$ has entries
$\partial\theta_m/\partial x_0 = -(y_0 - y_m)/r_m^2$ and
$\partial\theta_m/\partial y_0 = (x_0 - x_m)/r_m^2$,
with $r_m = \|\vect{p}_0 - \vect{p}_m\|$.
The localization CRB is $\text{CRB}(\vect{p}_0) = \tr[\FIM^{-1}(\vect{p}_0)]$ in $[\text{m}^2]$.

The Jacobian $\mat{J}$ has $K = M_t + M_r$ rows from diverse directions, which can improve the conditioning of $\FIM(\vect{p}_0)$ when informative directions are sufficiently separated.

Moreover, from~\eqref{eq:phase_rate}, the FIM-induced shift $\zeta_{m,n}\sin\theta_m$ depends on the local angle $\theta_m$, enabling each AP to independently tune its FIM shape to its own observation direction.

A closed form follows when the angular information matrix is diagonal. Sufficient conditions are orthogonal inter-AP waveforms, $\mat{R}_{m_t,m_t'}=\vect{0}$ for $m_t\neq m_t'$, together with $\mu_{m_r}=0$ at every Rx AP. Orthogonality alone does not remove the AoD--AoA cross terms. Under these conditions, $\FIM(\bm{\theta})$ has diagonal
entries $I_m \triangleq [\FIM(\bm{\theta})]_{m,m}$. Since each Jacobian row takes the form $\mat{J}_{m,:} = r_m^{-1}[-\sin\theta_m,\,\cos\theta_m]$, the $2\times 2$ position FIM becomes $\FIM(\vect{p}_0) = \sum_m (I_m/r_m^2)\,\vect{u}_m\vect{u}_m^\top$ with $\vect{u}_m = [-\sin\theta_m,\,\cos\theta_m]^\top$, and the localization CRB reduces to
\begin{equation}
    \mathrm{CRB}(\vect{p}_0) =
\frac{\sum_m I_m/r_m^2}
{\sum_{m<m'} \frac{I_m I_{m'}}{r_m^2 r_{m'}^2}\,\sin^2(\theta_m-\theta_{m'})},
\label{eq:crb_closedform}
\end{equation}
where {$I_m$} is the per-AP angular Fisher information and the denominator must be positive. For coupled angular information, the full Jacobian transformation must be used. The $\sin^2(\theta_m - \theta_{m'})$ terms in the denominator show that more diverse observation angles directly lower the CRB. This leads to \textbf{angular diversity amplification}, where the localization information scales with both per-AP Fisher information and the pairwise angular separation between APs. For co-located APs with $\theta_m = \theta$ $\forall m$, the denominator vanishes, confirming that a single observation angle cannot resolve 2D position.

With $w_m=I_m/r_m^2$, the determinant of the diagonal-case position
FIM explicitly measures bearing diversity:
\begin{equation}
\det\FIM(\vect{p}_0)=\sum_{m<m'}w_mw_{m'}
\sin^2(\theta_m-\theta_{m'}),
\label{eq:geometry_det}
\end{equation}
while $\tr\FIM(\vect{p}_0)=\sum_m w_m$. Parallel bearings can add
information without resolving the weak localization direction. The
correlated-waveform simulations instead use the full FIM.

For the full (possibly nondiagonal) angle-information matrix
$\mat{H}_\theta\succ0$, the Jacobian gives the sufficient bound
\begin{equation}
\tr[(\mat{J}^\top\mat{H}_\theta\mat{J})^{-1}]
\leq\frac{2}{\lambda_{\min}(\mat{H}_\theta)
\lambda_{\min}(\mat{J}^\top\mat{J})},
\label{eq:spectral_bound}
\end{equation}
provided $\mat{J}$ has rank two. This sufficient bound links angular information to the conditioning
of distributed bearings; it does not replace the exact CRB.

\subsection{Extension to Multiple Targets}
\label{sec:multitarget}
Consider $Q$ labeled targets at positions $\vect{p}_q$, sharing the same transmitted waveforms and array shapes. Retaining the calibrated-gain assumption, the noise-free signal is
\begin{equation}
\mat{S}_{m_r}=\sum_{q=1}^{Q}\sum_{m_t=1}^{M_t}
\alpha_{m_t,m_r,q}\vect{a}_{m_r,q}\vect{a}_{m_t,q}^{H}\mat{X}_{m_t},
\label{eq:multi_signal}
\end{equation}
where $\vect{a}_{m,q}$ is the array response toward target $q$ and the gains are held fixed in differentiation. This model considers unresolved returns within the same delay--Doppler cell; additional resolved measurements would change the information structure.
Let $\bm{\vartheta}=[\bm{\theta}_1^\top,\ldots,\bm{\theta}_Q^\top]^\top$ and define $\mat{B}_{m_r,q,i}=\partial\mat{S}_{m_r}/\partial\theta_{q,i}$. Each derivative follows the single-target expressions using only target $q$'s contribution. The $(q,q')$ angular information block is
\begin{equation}
[\mat{H}_{qq'}]_{ij}=\sum_{m_r}\frac{2}{\varsigma_{m_r}^2}
\Real\!\left\{\tr\!\left(\mat{B}_{m_r,q,i}^{H}\mat{B}_{m_r,q',j}\right)\right\}.
\label{eq:multi_fim}
\end{equation}
Thus $\mat{H}\in\mathbb{R}^{QK\times QK}$ contains both within-target and cross-target information. Orthogonal transmissions across APs do not generally eliminate $\mat{H}_{qq'}$ for $q\neq q'$, because the targets reflect the same waveforms.

For $\vect{p}=[\vect{p}_1^\top,\ldots,\vect{p}_Q^\top]^\top$, define $\mat{J}_{\rm all}=\diag(\mat{J}_1,\ldots,\mat{J}_Q)$ using the Jacobian of each target. The joint position information is $\mat{F}=\mat{J}_{\rm all}^{\top}\mat{H}\mat{J}_{\rm all}\in\mathbb{R}^{2Q\times2Q}$. If $\mat{F}\succ0$, target $q$'s localization bound is
\begin{equation}
\mathrm{CRB}_q=\tr\!\left[
\left(\mat{F}_{qq}-\mat{F}_{q,-q}\mat{F}_{-q,-q}^{-1}\mat{F}_{-q,q}\right)^{-1}\right],
\label{eq:multi_crb}
\end{equation}
where $-q$ denotes all other target positions. The positive-semidefinite Schur-complement subtraction quantifies the loss from jointly estimating the other targets. It vanishes when their position-information cross blocks vanish. A singular joint information matrix precludes a finite ordinary joint CRB; angle separation or increased local sensitivity alone does not establish identifiability. These are local bounds for labeled targets, not a detection or target-association guarantee. The numerical example evaluates this bound using single-target-designed shapes; joint multi-target shape optimization is not considered.

The distinction between joint and oracle bounds is rigorous whenever
the relevant information blocks are positive definite. Writing
$\mat{C}=\mat{F}_{q,-q}$ and $\mat{D}=\mat{F}_{-q,-q}$ gives
\begin{equation}
\mat{F}_{qq}-\mat{C}\mat{D}^{-1}\mat{C}^\top
\preceq\mat{F}_{qq},\qquad
\mathrm{CRB}^{\rm joint}_q\geq\mathrm{CRB}^{\rm oracle}_q.
\label{eq:joint_oracle_gap}
\end{equation}
The oracle inverts $\mat{F}_{qq}$ as if other positions were known.
Hence the gap in Fig.~\ref{fig:results}(f) measures coupling, not
estimator suboptimality; a near-singular Schur complement can make
the joint bound arbitrarily large.

\section{CRB-Driven Shape Design}
\label{sec:opt}
Let $\vect{z}$ collect all Tx and Rx element displacements. For a fixed waveform covariance and nominal target position, we minimize the sample-average localization bound
\begin{equation}
\min_{\vect{z}\in\mathcal{Z}}\quad
f(\vect{z})=\frac{1}{B}\sum_{b=1}^{B}
\tr\!\left[\FIM_b(\vect{p}_0;\vect{z})^{-1}\right],
\label{eq:shape_design}
\end{equation}
where $\mathcal{Z}$ contains the elementwise displacement bounds and $b$ indexes calibrated path-gain realizations. The expectation is approximated over phase realizations at a fixed target position, not over unknown target locations. A singular position-information matrix is assigned infinite cost.

We use coordinate search on the complete objective, so changes in both the diagonal and cross-information terms are retained. For coordinate $z_i$ with bound $b_i$, the candidates are $\{-b_i,-b_i/2,0,b_i/2,b_i\}$ together with its current value. A candidate is accepted only if it reduces $f$. Retaining the current value makes the objective non-increasing, but does not establish continuous stationary-point or global optimality. Each sweep requires at most five candidate evaluations per displacement variable. Experiments use $B=16$, zero initialization, and at most three sweeps; larger morphing ranges are warm-started by the preceding design.

With $D=M_tN_t+M_rN_r$ adjustable coordinates, one sweep takes at most
$5D$ new objective evaluations if the current value is cached. Here,
$D=64$ and $B=16$; the three-sweep budget is therefore at most 960
new shape-objective evaluations, each averaging 16 information matrices.
Monotonic acceptance concerns the training realizations only, motivating
the independent evaluation phases.

\begin{algorithm}[t]
\caption{CRB-Driven Coordinate Search}
\label{alg:shape}
\begin{algorithmic}[1]
\STATE Initialize a feasible shape $\vect{z}$ with finite $f(\vect{z})$.
\FOR{each sweep}
  \FOR{each displacement coordinate $z_i$}
    \STATE Evaluate the five grid candidates and the current value using~\eqref{eq:shape_design}.
    \STATE Retain the candidate with the smallest objective; keep the current value on ties.
  \ENDFOR
  \STATE Stop if the relative objective decrease is below $10^{-7}$.
\ENDFOR
\STATE Return the shape and recorded objective values.
\end{algorithmic}
\end{algorithm}

For the communication example, the resulting shapes are held fixed and combined with maximum-ratio transmission (MRT). Each AP allocates a fraction $\eta$ of its power to independent isotropic sensing and divides the remainder equally among users. With independent unit-power data symbols, the stacked transmit covariance is
\begin{equation}
\bar{\mat{R}}=\sum_{k=1}^{K_c}\tilde{\vect{w}}_k\tilde{\vect{w}}_k^H
+\diag\!\left(\frac{\eta P_1}{N_t}\mat{I}_{N_t},\ldots,
\frac{\eta P_{M_t}}{N_t}\mat{I}_{N_t}\right),
\label{eq:total_covariance}
\end{equation}
where $\tilde{\vect{w}}_k$ stacks the per-AP MRT vectors
$\vect{w}_{k,m_t}=\sqrt{(1-\eta)P_{m_t}/K_c}\,
\vect{h}_{k,m_t}/\|\vect{h}_{k,m_t}\|$.
Both data and sensing covariance contribute to the sensing information through $L\bar{\mat{R}}$. The resulting rate--CRB samples evaluate a feasible power-allocation rule, not a globally optimized Pareto frontier or an SDR-based joint design.

\section{Numerical Results}
\label{sec:sim}

\begin{figure*}[t]
\centering
\includegraphics[width=\textwidth]{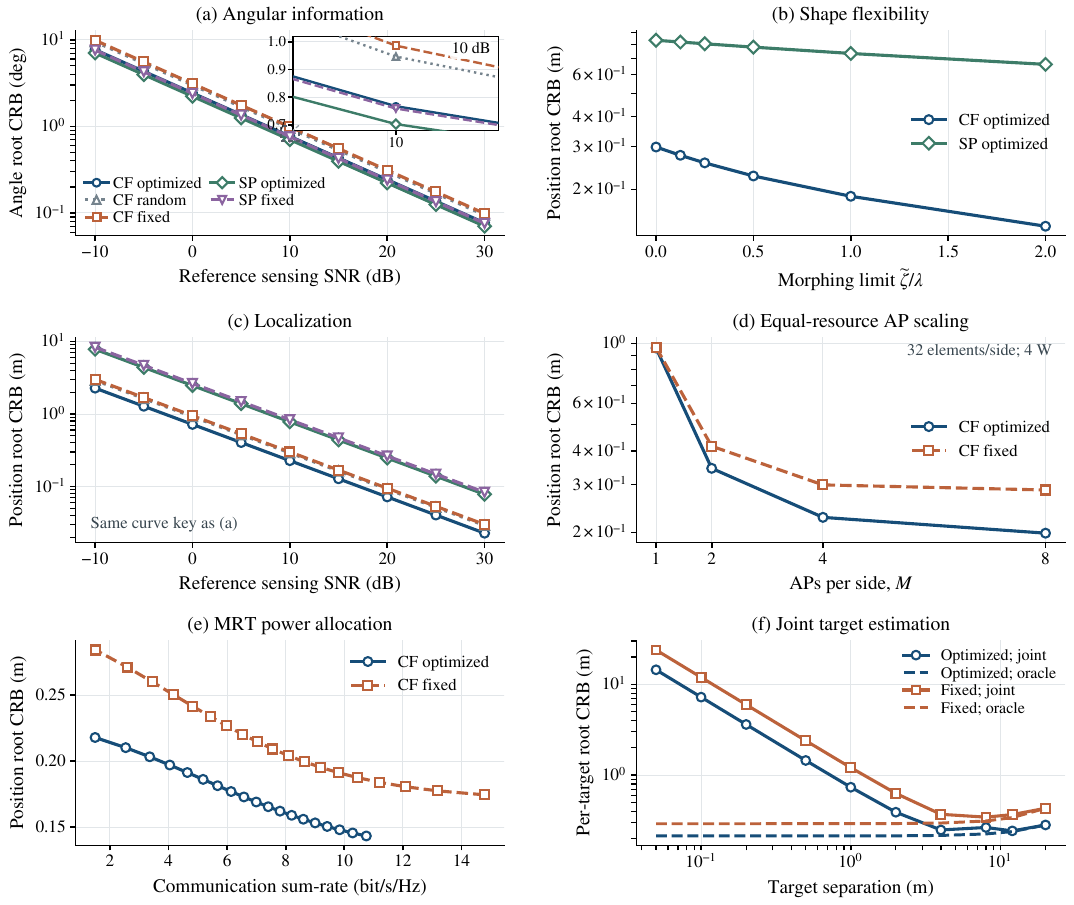}
\caption{Conditional bounds with calibrated gains. (a) Mean per-angle root CRB versus reference SNR, with a linear-scale inset magnifying the five 10-dB outcomes. (b) Position root CRB versus morphing limit at 10 dB. (c) Position root CRB versus reference SNR. (d) AP scaling with 32 elements per side and 4 W total transmit power. (e) MRT power-allocation samples using the complete transmit covariance. (f) Two-target joint and oracle position bounds; the oracle knows the other target's position. CF and SP denote cell-free and single-pair architectures. Panels (a)--(d) use 1000 held-out phase realizations; (e)--(f) use fixed realizations.}
\label{fig:results}
\end{figure*}

\subsection{Setup and Evaluation}
Four Tx and four Rx APs lie on a 40-m-radius circle centered at $(50,50)$ m, with the Rx ring rotated by $\pi/4$. The target is at $(66.437,46.333)$ m. Each AP has eight elements; the total transmit power is 4 W. The carrier frequency is 28 GHz, the observation length is $L=128$, and the default displacement limit is $0.5\lambda$. The single-pair baseline places 32-element Tx/Rx arrays at $(10,50)$ and $(90,10)$ m with the same total power. All steering vectors have unit norm.

Path-gain magnitudes are $1600/(r_{m_t}r_{m_r})$, with distances in metres and independent uniform phases known to the processor. Reference SNR is $(L/8)/\sigma^2$, corresponding to a unit-gain link with 1 W isotropic transmission over eight elements. All architectures use the same noise normalization. Sensing-only waveform Gram blocks are
$\mat{R}_{m_t,m_t'}=(LP_{\rm tot}/M_tN_t)c_{m_t,m_t'}\mat{I}_{N_t}$,
where $c_{m_t,m_t}=1$ and $c_{m_t,m_t'}=0.3$ otherwise. The full information matrix is used throughout; the diagonal-information closed form is not substituted into these correlated-waveform experiments.

Shapes are designed using 16 phase realizations and evaluated on 1000 independent held-out realizations. Random-shape baselines draw independent uniform displacements for each evaluation realization. Curves show $\sqrt{\mathbb{E}[\mathrm{CRB}]}$; the angle metric additionally divides the trace bound by $K$ before averaging. These are theoretical root bounds, not empirical estimator RMSE. Information singularities are not removed by diagonal loading.

To avoid an ambiguity in statistical reporting, the plotted quantity
is the square root of the phase-averaged conditional CRB, not the average
of individual square-root CRBs. The two operations are not equivalent
because the square-root function is nonlinear. The held-out phase bank
measures sensitivity of this ideal conditional information metric to the
assumed path-gain phases; it does not sample estimator errors or quantify
calibration mismatch. The communication and multi-target examples instead
use fixed stated realizations and should not be interpreted as averaged
reliability guarantees.

The reference-SNR sweep varies only the common sensing-noise scale,
without redesigning the shape. All information blocks then scale with
inverse noise power, so fixed-shape root CRBs vary with the inverse square
root of reference SNR. This explains the parallel trends in
Fig.~\ref{fig:results}(a) and (c), but need not hold for an
SNR-dependent waveform or geometry redesign.

\subsection{Single-Target Sensing}
Fig.~\ref{fig:results}(a)--(c) separates local angular accuracy from localization geometry. At 10 dB, cell-free fixed and CRB-designed arrays yield mean per-angle root CRBs of $0.986^\circ$ and $0.767^\circ$, and position root CRBs of 0.300 and 0.227 m. The latter reduction is approximately 2.40 dB. The random-shape position bound is 0.291 m. The single-pair fixed and designed bounds are 0.833 and 0.780 m, despite smaller angular bounds of $0.759^\circ$ and $0.703^\circ$. Thus angular accuracy alone does not determine localization performance. These gains are specific to the stated geometry and normalization.

The magnified 10-dB region in Fig.~\ref{fig:results}(a) makes the
five schemes distinguishable: their mean per-angle root CRBs are
$0.703^\circ$ (single-pair optimized), $0.759^\circ$ (single-pair fixed),
$0.767^\circ$ (cell-free optimized), $0.947^\circ$ (cell-free random),
and $0.986^\circ$ (cell-free fixed). In particular, the difference between
the single-pair fixed and cell-free optimized angular bounds is only
$0.008^\circ$. Nevertheless, the optimized cell-free architecture has a
much lower position root CRB than the optimized single pair. The distinction
is geometric: the location-domain Jacobian combines several distributed
bearings, whereas a single-pair arrangement can have accurate bearings
with less favorable intersection geometry.

Increasing the displacement limit improves the evaluated position bounds in Fig.~\ref{fig:results}(b). Panel (d) varies the AP count per side over $M\in\{1,2,4,8\}$ while fixing the total count at 32 elements per side and the total transmit power at 4 W; each AP therefore has $32/M$ elements. This comparison isolates distributed observation geometry under a fixed resource budget.

The optimized position root CRB falls from $0.961$ m at one AP per side
to $0.345$, $0.227$, and $0.199$ m at two, four, and eight APs, compared
with fixed-shape bounds of $0.968$, $0.416$, $0.300$, and $0.287$ m.
These resource-matched results reveal the gain from distributing a fixed
number of elements over more sites, with diminishing improvements in the
examined circular deployments.

\subsection{Communication and Multiple Targets}
For Fig.~\ref{fig:results}(e), users are at $(25,35)$ and $(75,70)$ m. Their channels are
$\vect{h}_{k,m_t}=(40/r_{k,m_t})e^{j\psi_{k,m_t}}\vect{a}_{m_t}(\theta_{k,m_t})$,
where $r_{k,m_t}$ is the user distance and the phases are fixed independent uniform draws. Communication noise variance is 0.01. We sweep $\eta$ from 0.05 to 0.95, maintaining 1 W per AP and evaluating the full communication interference and sensing covariance. In this realization, allocating more power to data transmission improves both rate and sensing because the data beams also illuminate the target. The curves therefore do not imply an unavoidable antagonistic trade-off.

Fig.~\ref{fig:results}(f) places two targets symmetrically along the horizontal axis about the nominal target. Both targets have equal calibrated gains, frozen at midpoint values to isolate separation effects, and occupy the same unresolved delay--Doppler cell. Single-target-designed shapes are reused. The joint position bound exceeds the oracle bound when the other position must also be estimated. At zero separation, equal gains make the joint information singular. The result illustrates information coupling, without claiming multi-target detection or association performance.

At $0.1$ m target separation, the optimized-shape joint and oracle
per-target root CRBs are $7.23$ and $0.214$ m; at $4$ m separation they
approach $0.249$ and $0.217$ m. This reduction in coupling does not imply
monotonic improvement at every separation, as both target bearings and
range-dependent information also change.

\section{Conclusion}
\label{sec:conclusion}
We derived conditional localization bounds for cell-free ISAC with fluid intelligent metasurfaces, retaining cross-AP information coupling and distinguishing angular sensitivity from observation geometry. The multi-target extension quantifies the cost of jointly estimating other target positions. A feasible coordinate-search design improves the localization bound in the evaluated geometry, while equal-resource comparisons reveal the value of distributed observations. Communication examples account for sensing information carried by data waveforms, and two-target results expose the effect of cross-target coupling. Future work will address unknown path gains, target-location uncertainty, and joint waveform--shape optimization.

\section*{Acknowledgment}
The authors acknowledge the support of the Communications and Computational Systems Laboratory (CCSL) at KAUST. The authors thank Jiguang He for correcting errors in algorithm in version1.

\bibliographystyle{IEEEtran}
\bibliography{IEEEabrv,references}

\end{document}